\documentclass[journal,twoside,web]{ieeecolor} 
\usepackage{tmi}
\usepackage{cite}
\usepackage{amsmath,amssymb,amsfonts}
\usepackage{algorithmic}
\usepackage{graphicx}
\usepackage[caption=false, font=footnotesize]{subfig}
\usepackage{lipsum}
\usepackage{textcomp}
\usepackage{verbatim}
\usepackage{booktabs,caption}
\usepackage{multirow}
\usepackage[T1]{fontenc}
\usepackage{algorithm}
\usepackage{algorithmic}
\usepackage{color}
\usepackage{siunitx}
\usepackage{threeparttable}
\usepackage{marvosym}
\usepackage{pifont} 
\usepackage{courier}  
\usepackage{listings}
\usepackage[colorlinks,hyperindex,breaklinks]{hyperref}
\newcommand{\ie}{\textit{i}.\textit{e}.}
\newcommand{\eg}{\textit{e}.\textit{g}.}
 
\newcommand{\cmark}{\ding{51}}

\newcommand{\myparagraph}[1]{\vspace{0.1em}\noindent\textbf{#1}}
\def\BibTeX{{\rm B\kern-.05em{\sc i\kern-.025em b}\kern-.08em
    T\kern-.1667em\lower.7ex\hbox{E}\kern-.125emX}}

\begin{document}
\title{Boosting Convolution with Efficient MLP-Permutation for Volumetric Medical Image Segmentation}
\author{Yi Lin,~\IEEEmembership{Graduate Student Member,~IEEE},
Xiao Fang,
Dong Zhang,~\IEEEmembership{Member,~IEEE}, \\
Kwang-Ting Cheng,~\IEEEmembership{Fellow,~IEEE},
and~Hao Chen,~\IEEEmembership{Senior Member,~IEEE}
\thanks{This work was supported by the Hong Kong Innovation and Technology Fund (Project No. ITS/028/21FP), Shenzhen Science and Technology Innovation Committee Fund (Project No. SGDX20210823103201011),  HKUST (Project No. FS111), Research Grants Council of the Hong Kong Special Administrative Region, China (Project No. T45-401/22-N), and the Project of Hetao Shenzhen-Hong Kong Science and Technology Innovation Cooperation Zone (HZQB-KCZYB-2020083).}
\thanks{Y. Lin, X. Fang, D. Zhang, K.-T. Cheng, and H. Chen are with the Department of Computer Science and Engineering, H. Chen is with the Department of Computer Science and Engineering and Department of Chemical and Biological Engineering, The Hong Kong University of Science and Technology, Hong Kong, China. 
H. Chen is also affiliated with HKUST Shenzhen-Hong Kong Collaborative Innovation Research Institute, Futian, Shenzhen, China. 
E-mail: \{yi.lin,~xfangal\}@connect.ust.hk;
\{dongz,~timcheng\}@ust.hk; 
jhc@cse.ust.hk.}
\thanks{Y. Lin and X. Fang contributed equally to this work.}
\thanks{H. Chen is the corresponding author.}}
\maketitle
\begin{abstract}
Recently, the advent of Vision Transformer (ViT) has brought substantial advancements in 3D benchmarks, particularly in 3D volumetric medical image segmentation (Vol-MedSeg).
Concurrently, multi-layer perceptron (MLP) network has regained popularity among researchers due to their comparable results to ViT, albeit with the exclusion of the resource-intensive self-attention module.
In this work, we propose a novel permutable hybrid network for Vol-MedSeg, named PHNet, which capitalizes on the strengths of both convolution neural networks (CNNs) and MLP.
PHNet addresses the intrinsic anisotropy problem of 3D volumetric data by employing a combination of 2D and 3D CNNs to extract local features.
Besides, we propose an efficient multi-layer permute perceptron (MLPP) module that captures long-range dependence while preserving positional information.
This is achieved through an axis decomposition operation that permutes the input tensor along different axes, thereby enabling the separate encoding of the positional information.
Furthermore, MLPP tackles the resolution sensitivity issue of MLP in Vol-MedSeg with a token segmentation operation, which divides the feature into smaller tokens and processes them individually.
Extensive experimental results validate that PHNet outperformed the state-of-the-art methods with lower computational costs on the widely-used yet challenging COVID-19-20, Synapse, LiTS and MSD BraTS benchmarks.
The ablation study also demonstrated the effectiveness of PHNet in harnessing the strengths of both CNNs and MLP. The code is available on Github: \href{https://github.com/xiaofang007/PHNet}{https://github.com/xiaofang007/PHNet}.
\end{abstract}
\section{Introduction}
\label{sec:introduction}
Medical image segmentation (MedSeg) is a fundamental yet challenging task in medical image analysis~\cite{minaee2021image}. The primary objective of MedSeg lies in the precise identification and localization of semantic lesions or anatomical structures within medical images obtained through imaging modalities such as Computed Tomography, X-ray, and Magnetic Resonance Imaging~\cite{isensee2021nnunet,zhang2022deep,lin2023rethinking,chen2023confidence}. 
It serves as an important auxiliary tool in various clinical applications, such as disease diagnosis~\cite{shi2020review}, treatment planning~\cite{wang2021multiclass}, and monitoring~\cite{liu2022region}.
Over the past decades, substantial research efforts have focused on developing efficient and robust MedSeg methodologies.
One of the most popular architectures for this task is UNet~\cite{ronneberger2015u}, which employs an encoder-decoder structure and skip connections to reserve both contextual and semantic information. Building upon the success of UNet, numerous variants have been proposed with various convolution-based blocks and different skip connections strategies, including ResUNet~\cite{he2016deep}, Y-Net~\cite{mehta2018net}, and V-Net~\cite{milletari2016v}, etc.

\begin{figure}[t]
\centering
\includegraphics[width=0.48\textwidth]{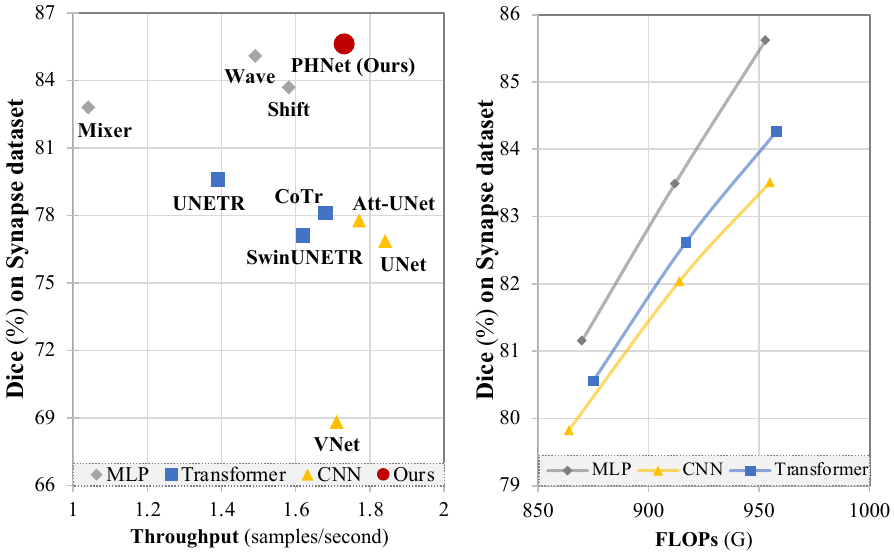}
\caption{Left: Performance \textit{v.s.} throughput of PHNet and other SOTA methods.  Right: performance \textit{v.s.} FLOPs of CNN, Transformers, and MLP, in different model capacity.}
\label{fig:throughput}
\end{figure}

Recently, Transformers with attention mechanism have shown promising superiority in the realm of natural language processing~\cite{vaswani2017attention}. 
Subsequent studies, such as ViT~\cite{dosovitskiy2020vit} and DeiT~\cite{touvron2020training}, have demonstrated remarkable capabilities in achieving state-of-the-art performance on versatile computer vision tasks.
Given the notable strides of Transformers in natural image recognition tasks, researchers have investigated the effectiveness of these neural networks for MedSeg.
To name a few, TransUNet~\cite{chen2021transunet} proposed to employ a Transformer in the bottleneck of a UNet architecture for global information communication.
Similarly, UNETR~\cite{hatamizadeh2022unetr}, CoTr~\cite{xie2021cotr}, and SwinUNet~\cite{cao2021swinunet} designed a hierarchical fusion of Transformer and CNNs architecture.

Despite the developments in MedSeg domain via Transformer methods, the computational complexity of self-attention increases quadratically with the size of input images.
The heavy computational costs of Transformer methods limit their practical application, particularly for 3D volumetric medical images, especially in high-resolution images with dense features across scales that necessitate a substantial number of forward and backward passes~\cite{valanarasu2022unext}.
Consequently, multi-layer perceptron (MLP) has regained interest in the community, as it has demonstrated comparable performance with both CNNs and Transformers (in Figure~\ref{fig:throughput}), without requiring the heavy self-attention mechanism~\cite{tolstikhin2021mixer}.
However, the effectiveness of MLP in volumetric MedSeg remains understudied. 

In light of this, we investigate the effectiveness and efficacy of CNN, Transformer, and MLP for volumetric MedSeg.
We analyze the main challenges of volumetric MedSeg.
\textbf{First}, the critical challenge is to capture the long-range dependencies in volumetric medical images while maintaining the local contextual information.
\textbf{Second}, the inherent anisotropy problem of volumetric medical images, as depicted in Figure~\ref{fig:anisotropic}, requires the network to capture the varying information density in different directions.
\textbf{Third}, an affordable computational complexity for volumetric MedSeg holds paramount importance to ensure practical applicability in real-world scenarios.

\begin{figure}[t]
\centering
\includegraphics[width=0.48\textwidth]{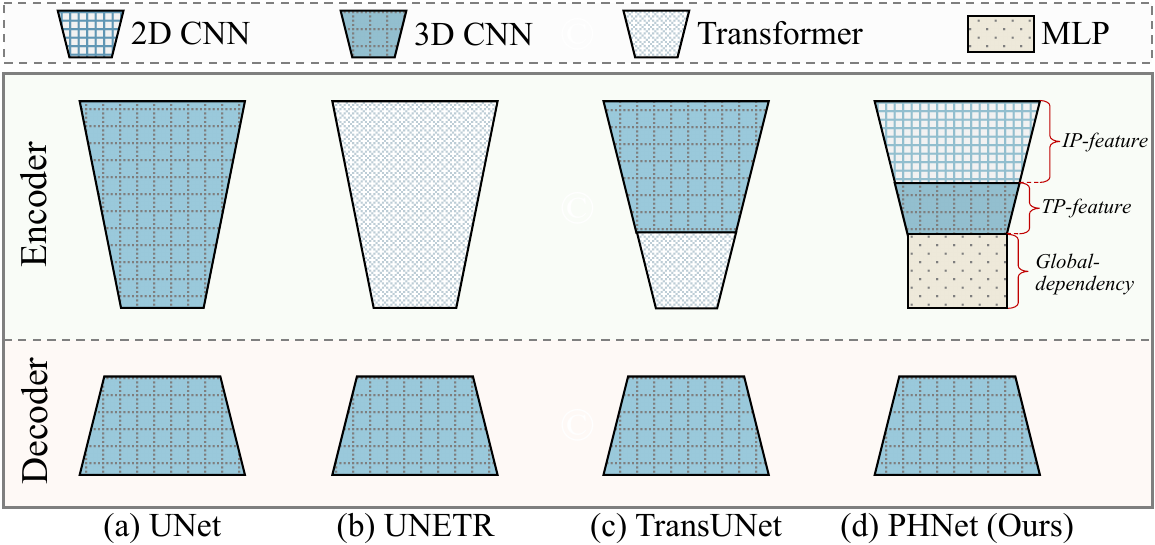}
\caption{Overview of the proposed PHNet.}
\label{fig:overview}
\end{figure}

To address aforementioned challenges, in this paper, we propose PHNet, a novel \textbf{P}ermutable \textbf{H}ybrid \textbf{Net}work that enjoys the strengths of CNN and MLP for volumetric medical image segmentation. 
The CNN is employed to capture the local context information, while the MLP is utilized to capture the long-range dependencies in volumetric medical images.
As illustrated in Figure~\ref{fig:overview}, PHNet embodies an encoder-decoder paradigm. 
Notably, the encoder utilizes a 2.5D CNN structure that capitalizes on the inherent anisotropy of medical images, while avoiding information loss in shallow layers by capturing the varying information density in different directions of volumetric medical images. 
In PHNet, we further propose MLPP, a {M}ulti-Layer {P}ermute {P}erceptron module that can maintain the positional information with the axial decomposition operation while integrating global interdependence in a computationally efficient manner. To enhance computational efficiency, token-group operation is introduced, which efficiently aggregates feature maps at a token level, reducing the number of computations required. PHNet is evaluated on four publicly available datasets, COVID-19 Lung CT Lesion Segmentation Challenge-2020~\cite{roth2022rapid}, Synapse Multi-Organ Segmentation~\cite{landman2015miccai}, Liver Tumor Segmentation~\cite{bilic2023liver} and the Medical Segmentation Decathlon brain tumor segmentation~\cite{antonelli2022medical}.
Extensive experimental results validate that PHNet achieves state-of-the-art performance on both datasets, surpassing the winner in the MICCAI COVID-19-20 challenge.
In summary, the contributions of our study are
as follows:
\begin{itemize}
\item We theoretically and empirically analyze the effectiveness and efficiency of three key neural network architectures (CNN, Transformer, and MLP) for volumetric MedSeg.
\item We propose PHNet, a novel permutable hybrid network, which is the first attempt to integrate CNN and MLP for
volumetric MedSeg.
\item We design an MLP permutation module,
namely MLPP, to enhance segmentation performance and improve efficiency through axial decomposition and token segmentation operations.
\item We conduct extensive experiments on four public datasets and the results demonstrate the effectiveness of PHNet. 
\end{itemize}
\section{Related Work}
\label{sec:related}
\myparagraph{CNN-Based Networks.} 
Since the groundbreaking introduction of the UNet~\cite{ronneberger2015u}, CNN-based networks have achieved state-of-the-art results on various 2D and 3D MedSeg tasks~\cite{dou20163d,falk2019unet,dong2022mnet,zhang2020causal}.
These methods use CNNs as the backbone to extract image features, and combine some elaborate tricks (\eg, skip connection, multi-scale representation, feature interaction) for feature enhancement. 
Compared to 2D methods, 3D approaches directly utilize the full volumetric image represented by a sequence of 2D slices or modalities.
Despite their success, CNN-based networks exhibit a constraint in effectively learning global context and long-range spatial dependencies, resulting in sub-optimal performance for challenging tasks.

\myparagraph{Transformer-Based Networks.} 
Vision Transformers have been applied in MedSeg to establish long-range dependence and capture context information. 
Existing methods can be categorized into two groups: i) pure Transformers and ii) hybrid architectures that combine CNNs and Transformers. 
In the first category, CNNs in a UNet-like encoder-decoder architecture are replaced with pure Transformer blocks. 
For instance, SwinUNet~\cite{cao2021swinunet} and MissFormer~\cite{huang2022missformer} utilize Swin Transformer blocks~\cite{liu2021swin} and Enhanced Transformer blocks, respectively. 
In the second category, researchers strive to harness the benefits of both CNNs and ViT by capturing local information and long-range feature dependencies simultaneously. 
For example, UNETR~\cite{hatamizadeh2022unetr} employs Transformers as encoders to learn sequence representations, enabling the capture of global multi-scale information, while CNNs serve as decoders to enhance localization ability. CTO-Net~\cite{lin2023rethinking} follows the standard encoder-decoder segmentation paradigm. The
encoder network incorporates Res2Net for capturing
local semantic information, and a lightweight ViT for integrating
long-range dependencies. To enhance the learning capacity on boundary, a
boundary-guided decoder network is proposed that uses a boundary mask
obtained from Sobel operators as explicit supervision.
To address the computational costs associated with these architectures, various approaches have been proposed.
Peiris~\textit{et al.} introduce VT-UNet~\cite{peiris2022robust}, which leverages a hierarchical vision Transformer that gradually decreases feature resolution in Transformer layers and incorporates sub-sampled attention modules. 
Chen~\textit{et al.} propose TransUNet~\cite{chen2021transunet}, which uses CNN to efficiently extract local information in shallow layers and employs a Transformer to model global features in the bottleneck layer. 
Tang~\textit{et.al} propose SwinUNETR~\cite{tang2022swinunetr} that calculates the attention weight between different tokens in shifted local windows, thereby reducing the computational cost from quadratic to linear complexity. 
However, it is worth noting that the self-attention mechanism remains computationally demanding and relatively slow when executed on GPUs~\cite{lee20233d,zhang2020feature}.

\myparagraph{MLP-Based Networks.} 
Acknowledging the substantial computational cost of attention blocks in Transformer, simple and efficient modules that consist of only MLPs are proposed. 
Notably, MLP-Mixer~\cite{tolstikhin2021mixer} uses token-mixing MLP and channel-mixing MLP to capture the relationship between tokens and channels, respectively.
gMLP~\cite{liu2021gmlp} splits the input feature maps into two separate parts for channel projection and spatial projection. Then, a gating unit is used to merge the two branches. 
However, these methods encounter two main challenges.
First, the direct flattening of input feature maps into 1D vectors for enabling cross-token interactions results in the loss of positional information carried by 2D input feature maps. 
Additionally, this flattening operation escalates the computational complexity significantly, with a quadratic increase relative to the size of the input feature maps.
Second, these architectures suffer from image resolution sensitivity issue~\cite{liu2022mlpsurvey}, which means they cannot handle input with flexible resolutions, limiting their adaptability. 
To address these challenges, several methods have been proposed.
CycleMLP~\cite{chen2022cyclemlp} introduces a hierarchical cycle fully connected layer to aggregate spatial context and improve performance. 
WaveMLP~\cite{tang2022wave} represents each token as a wave function with amplitude and phase to dynamically aggregate features according to the semantic contents of input images. 
In MedSeg field, UNeXt~\cite{valanarasu2022unext} shifts tokens along vertical and horizontal directions to get an axial receptive field of 2D images to increase efficiency. 
However, the effects of MLPs in Vol-MedSeg remain unexplored.
To the best of our knowledge, it is the first attempt to investigate the effectiveness of integrating CNNs and MLPs in Vol-MedSeg.
\section{Methodology}
This section describes the overall pipeline of PHNet first, and then shows the details of the 2.5D convolutional network, Multi-Layer Permute Perceptron module and decoder network.

\subsection{Overall pipeline} 
\begin{figure}[htbp]
    \centering
    \includegraphics[width=0.48\textwidth]{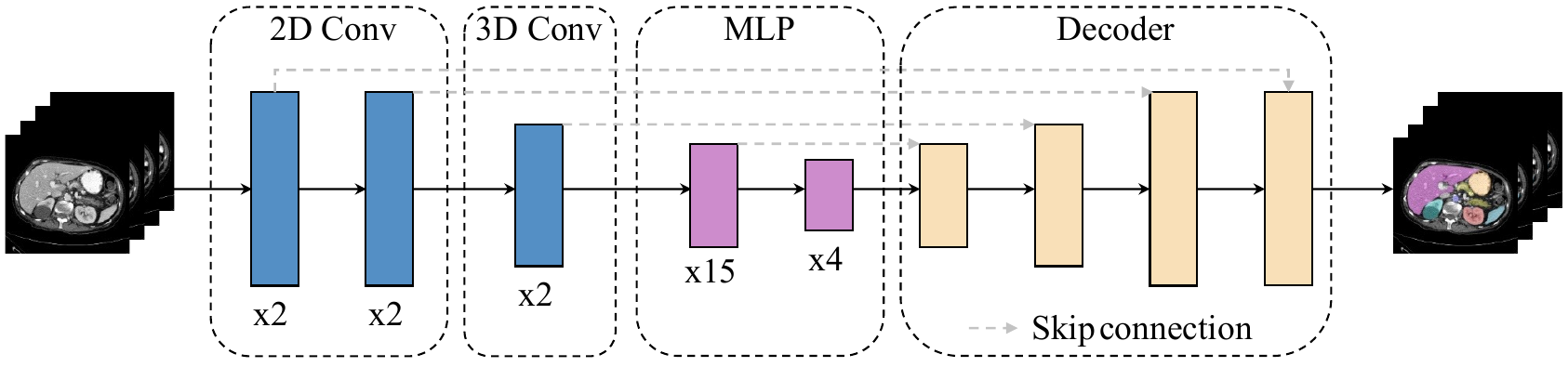}
    \caption{Detailed network architecture of PHNet.}
    \label{fig:framework}
\end{figure}
The network architecture, as illustrated in Figure~\ref{fig:framework}, can be segmented into several stages. 
The initial stage involves a 2.5D convolutional network, where the input images undergo two consecutive 2D conv-blocks and 2D Max-pooling operations. 
This sequence is thoughtfully designed to prioritize the exploration of in-plane features during training, with the goal of achieving a more harmonized resolution across all dimensions. 
Following this, a 3D conv-block is employed to further consolidate local features in the shallow layers.

After the convolutional stage, the resultant output features are passed to the MLP stage, which is meticulously crafted to capture long-range dependencies. 
The MLP stage encompasses two layers, with the first layer containing fifteen Multi-Layer Permute Perceptron (MLPP) blocks and the second layer consisting of four MLPP blocks. 
This stage comprises multiple Multi-Layer Permute Perceptron (MLPP) blocks, which have been designed to facilitate the deep aggregation of global features, while maintaining a focus on computational efficiency.

Finally, the output features from each layer are fed into a CNN-based decoder, which progressively upsamples the feature maps to match the resolution of the input images.

\begin{figure}[thbp]
\centering
\includegraphics[width=0.45\textwidth]{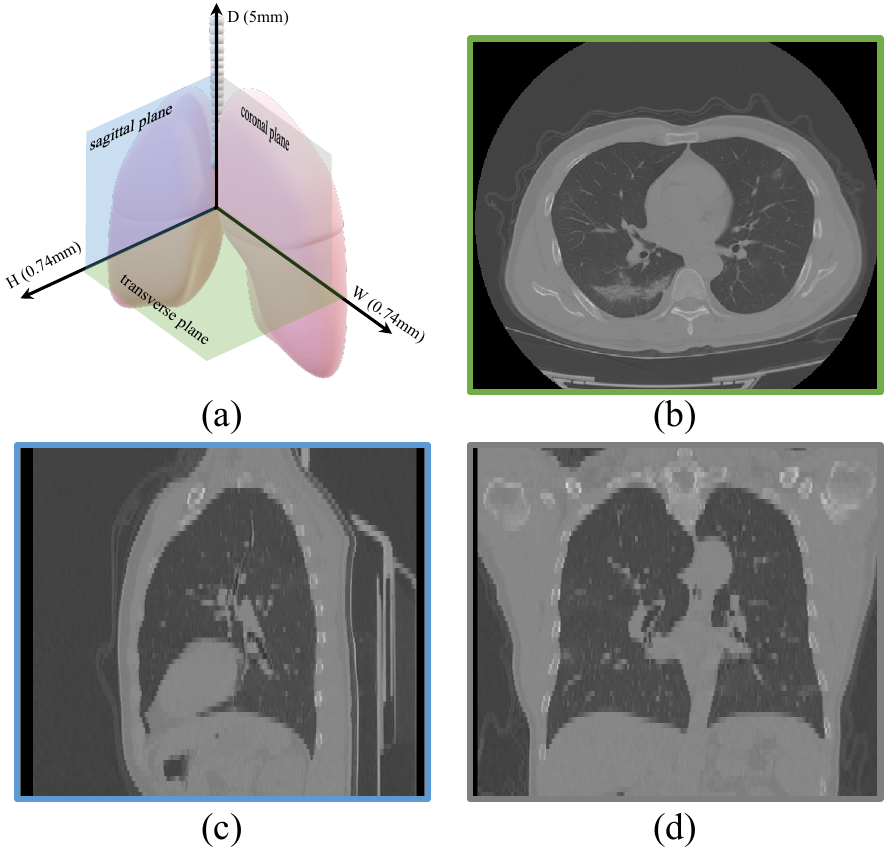}
\caption{Illustration of the anisotropic problem. (a) Anatomical planes of Lung including (b) transverse plane (c) sagittal plane (d) coronal plane. Better visualization with zooming in.}
\label{fig:anisotropic}
\end{figure}
\begin{figure*}[!t]
\centering
\includegraphics[width=\textwidth]{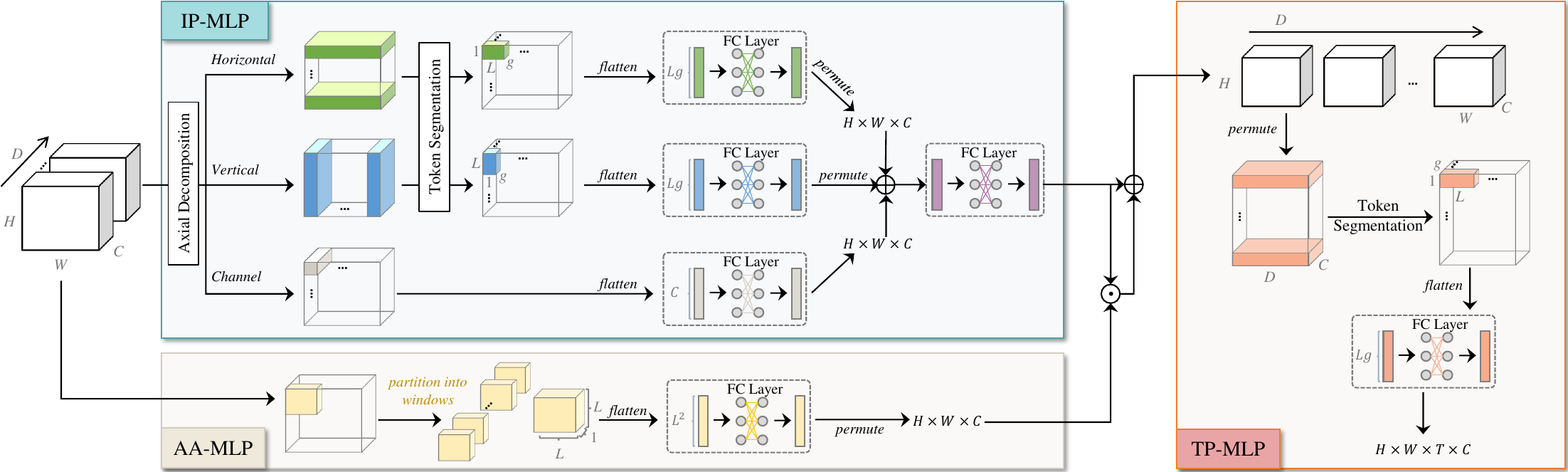}
\caption{Illustration of multi-layer permute perceptron (MLPP) module.}
\label{fig:mlpp}
\end{figure*}

\subsection{2.5D Convolution} 
\label{sec2.1}
Inspired by previous research in medical image segmentation~\cite{battaglia2018bias} and the anisotropic nature of volumetric medical images, we incorporate convolutional layers in the shallow layers of the encoder to extract local features.
Volumetric images, such as CT and MRI scans, are often affected by anisotropic problems due to their thick-slice scanning~\cite{wang20192d5,liu20183d}, resulting in high in-plane (IP) resolution and low through-plane (TP) resolution~\cite{dong2022mnet}.
This discrepancy in information density between the depth (D) axis and the in-plane (H and W) dimensions is a prominent observation. 
The D-axis exhibits significantly larger spacing between slices, resulting in a relatively sparse representation of information. In contrast, the in-plane dimensions maintain a higher resolution, preserving a denser amount of information.
To illustrate this issue, we refer to an example from the COVID-19-20 dataset~\cite{roth2022rapid}. 
As depicted in Figure~\ref{fig:anisotropic}(a), the average in-plane (IP) resolution is 0.74mm, considerably higher than the through-plane (TP) resolution of 5mm. 
This discrepancy is further highlighted in Figure~\ref{fig:anisotropic}(b-d), where sagittal and coronal plane images appear blurrier compared to the transverse plane image. The application of isotropic kernels directly to learn features in such scenarios poses a challenge. 
The highly discontinuous and sparse information along the D-axis introduces long-range noise to the representation of local features within the transverse plane~\cite{zeng2017deepem3d,li2020z}.
This noise can increase the network's vulnerability to overfitting and hinder optimal generalization.
To mitigate this issue, we use 2D conv-blocks to capture the IP information until the feature is reformulated in approximately uniform resolution across all three axes. Then, we apply 3D conv-blocks to handle the volumetric information. 
Each encoder layer consists of two residual convolution blocks, with each block comprising two sequential Convolution--Instance Normalization--ReLU operations~\cite{ulyanov2016instancenorm}.
The residual addition takes place before the final ReLU activation.

\subsection{Multi-Layer Permute Perceptron}
\label{sec2.2}
Although CNNs are capable of modeling long-range dependencies through deep stacks of convolution layers, studies~\cite{valanarasu2022unext,hou2022vip,tolstikhin2021mixer,zhang2022morphmlp} have highlighted the superior ability of MLP-based networks to learn global context. 
Motivated by this, we design MLPP as depicted in Figure~\ref{fig:mlpp} to acquire global information in deep layers of the encoder.
MLPP decomposes the training of IP feature and TP feature in sequential order. 
We denote these two blocks as \textbf{IP-MLP} and \textbf{TP-MLP}, respectively.
To facilitate the communication of cross-axis tokens, we further propose an auxiliary attention branch in IP-MLP, denoted as \textbf{AA-MLP}.
This enlarged receptive field facilitates the integration of information from distant tokens and improves the model's ability to capture global dependencies while maintaining awareness of the relative positions of elements.

\subsubsection{\emph{IP-MLP}.} 
Given one slice of input feature maps $\mathbf{X}\in \mathbb{R}^{H\times W \times C}$, 
as depicted in Figure~\ref{fig:mlpp}, in contrast to conventional MLP-based methods~\cite{liu2021gmlp,tolstikhin2021mixer} that lose the spatial information of the original conv-features and require large computational costs, we introduce an axial decomposition operation in triplet pathways that separately process $\mathbf{X}$ in horizontal ($W$), vertical ($H$), and channel ($C$) axis. 
It is achieved by rearranging $\mathbf{X}$ to the desired order such that voxels within the same pathway across all channels are grouped, thus enabling the preservation of precise positional information along other axes when encoding information along one axis.
Additionally, this approach effectively reduces computational complexity, which scales linearly with the size of $\mathbf{X}$.

To strike a balance between local feature and long-distance interactions, and alleviate image resolution sensitivity problem~\cite{liu2022mlpsurvey} wherein the MLP-based model is sensitive to the input resolution,
we present a token segmentation operation that splits the feature vector into multiple tokens, which can be efficiently processed by the following fully-connected (FC) layers.
We take the horizontal axis as an example. 
Instead of encoding the entire dimension, we split $\mathbf{X}$ into non-overlapping segments along the horizontal direction.
Each segment, denoted as $\mathbf{X}{i}\in \mathbb{R}^{L\times C}$, where $i \in {1,...,HW/L}$, has a segment length of $L$. 
Similarly, we divide each $\mathbf{X}_{i}$ into multiple non-overlapping groups along channel dimension, where each group has $g = C/L$ channels. 
This yields split segments, and each individual segment is $\mathbf{X}_{i}^{k}\in \mathbb{R}^{Lg}$, where $ k \in \{1,...,C/g\}$. 
Next, we flatten each segment and map $\mathbb{R}^{Lg} \mapsto \mathbb{R}^{Lg}$ by an FC layer to transform each segment, producing $\mathbf{Y}_{i}^{k}$. 
To recover the original dimension, we permute all segments back to $\mathbf{Y}_{W} \in \mathbb{R}^{H\times W \times C}$. 
Similarly, we conduct the same operations in the vertical pathway as above to permute tokens along the vertical direction, yielding $\mathbf{Y}_{H}$. 
To facilitate the communication among groups along channel dimension, we design a parallel branch that contains an FC layer that maps $\mathbb{R}^{C} \mapsto \mathbb{R}^{C}$ to process each token individually, yielding $\mathbf{Y}_{C}$. 
Finally, we feed the element-wise summation of horizontal, vertical, and channel features into a new FC layer to attain the output, which can be formulated as:
\begin{equation} \label{eqn:fusion2}
    \mathbf{Y}_{\text{IP}} = (\mathbf{Y}_H + \mathbf{Y}_W + \mathbf{Y}_C)\mathbf{W},
\end{equation}
where $\mathbf{W} \in \mathbb{R}^{C\times C} $ denotes an FC weight matrix. 
In this way, the number of parameters in MLPP does not rely on the shape of the input feature maps, and thus MLPP can accommodate a more flexible input shape. We also provide an example of the code based on the PyTorch as shown in Algorithm~\ref{alg:IP_MLP}.

\begin{algorithm}[t]
\caption{Pseudo code for IP-MLP (PyTorch-like)}
\label{alg:IP_MLP}
\definecolor{codeblue}{rgb}{0.25, 0.5, 0.25}
\lstset{
backgroundcolor=\color{white}, 
basicstyle= \fontsize{8pt}{8pt}\ttfamily\selectfont, 
columns=fullflexible, 
breaklines=true, 
captionpos=b, 
commentstyle=\fontsize{6.5pt}{6.5pt}\color{codeblue}, 
keywordstyle=\fontsize{6.5pt}{6.5pt}\color{blue}
}
\begin{lstlisting}[language=python]
# x: input tensor of shape (H, W, D, C)
# L: segment length

def init():
    g = C // L
    proj_h = Linear(C, C)    # vertical axis.
    proj_w = Linear(C, C)    # horizontal axis.
    proj_c = Linear(C, C)    # channel information.
    proj_1 = Linear(C, C) # information fusion. 

def IP-MLP(x): 
    # Encode vertical axis information.
    x_h = x.transpose(1, 0)
    x_h = x_h.reshape(H*W//L, L, D, L, g)
    x_h = x_h.permute(0, 3, 2, 1, 4)
    x_h = x_h.reshape(H*W//L, L, D, L*g)
    x_h = proj_h(x_h)
    x_h = x_h.reshape(H*W//L, L, D, L, g)
    x_h = x_h.permute(0, 3, 2, 1, 4)
    x_h = x_h.reshape(W, H, D, C)
    x_h = x_h.transpose(1, 0)
    # Encode horizontal axis information.
    x_w = x.reshape(H*W//L, L, D, L, g)
    x_w = x_w.permute(0, 3, 2, 1, 4)
    x_w = x_w.reshape(H*W//L, L, D, L*g)
    x_w = proj_w(x_w)
    x_w = x_w.reshape(H*W//L, L, D, L, g)
    x_w = x_w.permute(0, 3, 2, 1, 4)
    x_w = x_w.reshape(H, W, D, C)
    # Encode channel information.
    x_c = proj_c(x)
    # Fusion
    x = x_h + x_w + x_c
    x = proj_1(x)
    return x
    
\end{lstlisting}
\end{algorithm}

\subsubsection{\emph{AA-MLP}.} 

\begin{figure}[htbp]
\centering
\includegraphics[width=0.48\textwidth]{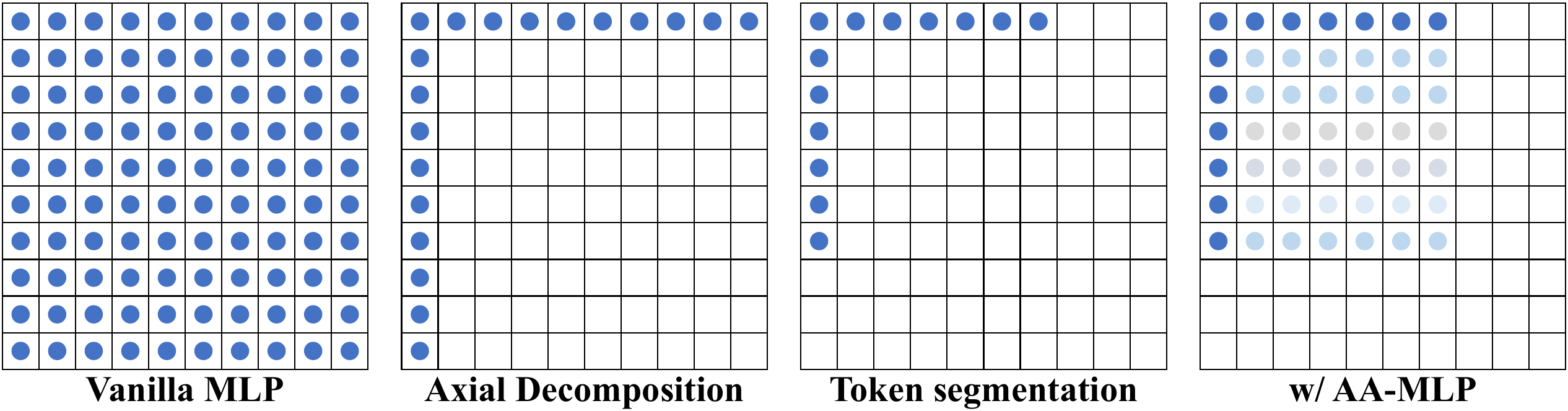}
\caption{Sampling locations of each operation.}
\label{fig:receptive_field}
\end{figure}
The IP-MLP, despite its strengths, possesses two limitations that may have a negative impact on segmentation performance, as depicted in Figure~\ref{fig:receptive_field}. First, the axial decomposition operation restricts direct interactions among tokens that are not in the same horizontal or vertical position. Second, the token segmentation operation would suffer from a smaller local reception field compared with the vanilla MLP~\cite{tolstikhin2021mixer}. To overcome these limitations, we design an auxiliary parallel branch to enable intra-axis token communication, which serves as an attention mechanism via a lightweight yet effective MLP-like architecture.
Specifically, given an input slice of feature maps $\mathbf{X}\in \mathbb{R}^{H\times W \times C}$, we partition $\mathbf{X}$ into non-overlapping windows. 
We set window size to $L$ and thus obtain $\mathbf{X}_{i}\in \mathbb{R}^{L\times L}$, where $i \in \{1,...,HWC/L^{2}\}$. 
Subsequently, we apply an FC matrix $\mathbf{W}\in \mathbb{R}^{L^{2}\times L^{2} }$ to transform each window and get $\mathbf{Y}_{i} \in \mathbb{R}^{L\times L}$. 
The final attention map $\mathbf{Y}_A \in \mathbb{R}^{H\times W\times C} $ is obtained by permuting all windows back to the original dimension. 
Finally, the feature maps $\mathbf{F}_{\text{IP}}$ of IP-MLP are obtained by performing residual attention~\cite{wang2017residual} of $\mathbf{Y}_{\text{IP}}$ and $\mathbf{Y}_A$ as follows:
\begin{equation} \label{eqn:fusion}
    \mathbf{F}_{\text{IP}} = (1+\mathbf{Y}_A) \odot \mathbf{Y}_{\text{IP}},
\end{equation}
where $\odot$ denotes element-wise multiplication.
 
\subsubsection{\emph{TP-MLP}.} 
Upon obtaining the in-plane information from the IP-MLP and AA-MLP module, we proceed to apply the TP-MLP module to capture the long-term through-plane features. Similarly, given feature maps $\mathbf{F}_{\text{IP}}\in \mathbb{R}^{H\times W\times D \times C}$, which is a combination of output feature maps from IP-MLP and AA-MLP, we first split $\mathbf{X}=\mathbf{F}_\text{IP}$ along the depth dimension into non-overlapping segments with a segment length of $L$.
We thus obtain $\mathbf{X}_{i}\in \mathbb{R}^{L\times C}$, where $ i \in \{1,...,HWD/L\}$. 
Subsequently, we divide $\mathbf{X}$ into several non-overlapping groups along the channel dimension, with each group containing $g = C/L$ channels. 
This results in $\mathbf{X}_{i}^{k}\in \mathbb{R}^{Lg}$, where $ k \in \{1,...,C/g\}$. 
Subsequently, we flatten each segment and map $\mathbb{R}^{Lg} \mapsto \mathbb{R}^{Lg}$ by an FC layer, yielding $\mathbf{Y}_{i}^{k}$. 
Finally, we permute all segments $\mathbf{Y}_{i}^{k} \in \mathbb{R}^{Lg} $ back to the original dimension, yielding the output $\mathbf{F}_{\text{TP}} \in \mathbb{R}^{H\times W\times D \times C}$.

\subsection{Decoder Network}
The decoder in our proposed method utilizes a slim CNN architecture, employing transpose convolution to progressively upsample the feature maps to match the input image resolution. 
Within the decoder, we separate an isotropic 3D convolution with kernel size $3\times3\times3$ into a $3\times3\times1$ in-plane convolution and a $1\times1\times3$ through-plane convolution to efficiently fuse the feature~\cite{zhang2021efficient}. 
We further include skip connections between the encoder and decoder, allowing for the preservation of low-level details. 
\section{Experiments}
\subsection{Datasets and Evaluation Metrics}
We conduct experiments on four publicly available datasets: COVID-19-20~\cite{roth2022rapid}, Synapse~\cite{landman2015miccai}, LiTS~\cite{bilic2023liver} and MSD BraTS~\cite{antonelli2022medical}. 
COVID-19-20 is comprised of 249 unenhanced chest CT scans, with 199 samples designated for training and 50 samples for testing. All samples are positive for SARS-CoV-2 RT-PCR.
Synapse consists of 30 abdominal CT scans, with 14, 4, and 12 cases designated for training, validation, and testing, respectively~\cite{cao2021swinunet}. 
LiTS consists of 131 contrast-enhanced abdominal CT scans, with 90, 15, and 26 cases for training, validation, and testing~\cite{lin2023rethinking}. 
MSD BraTS consists of 484 MRI scans in isotropic resolutions from patients diagnosed with either glioblastoma or lower-grade glioma, each containing four channels: FLAIR, T1w, T1gd, and T2w. The scans include ground truth labels for three tumor sub-regions: edema (ED), enhancing tumor (ET), and non-enhancing tumor (NET).
For COVID-19-20, we use the official evaluation metrics from the challenge~\cite{roth2022rapid}, including Dice coefficient (Dice), Intersection over Union (IoU), Surface Dice coefficient (SD), Normalized Volume Difference (NVD), and Hausdorff Distance (HD). 
For LiTS, we follow~\cite{lin2023rethinking} to adopt Dice and IoU as evaluation metrics. For MSD BraTS and Synapse, we follow~\cite{peiris2022robust} and~\cite{chen2021transunet} to adopt Dice and HD as evaluation metrics.
The computational cost is measured with an input size of $192\times192\times48$ and a batch size of 1, in terms of FLOPS (G), Parameters (M), Peak Memory (G), and Throughput (samples/s).
For details, please refer to \cite{hou2022vip}. 

\begin{table*}[t]
    \centering
    \caption{Result comparisons with the state-of-the-art methods on Synapse~\cite{landman2015miccai}.} 
    \renewcommand\arraystretch{1.2}
    \setlength{\tabcolsep}{9pt}{
    \begin{tabular}{ c | r | c c | c c c c c c c c} 
    \toprule
    \multicolumn{2}{c|}{\textbf{Methods}} & \textbf{mDice} $\uparrow$ & \textbf{HD} $\downarrow$ & \textbf{Aorta} & \textbf{Gallb}. & \textbf{Kid(L)} & \textbf{Kid(R)} & \textbf{Liver} & \textbf{Panc}. & \textbf{Spleen} & \textbf{Stom}. \\
    \midrule
    \multirow{4}{*}{\rotatebox{90}{CNN-based}}& UNet~\cite{ronneberger2015u} & 76.85 & 39.70 & 89.07 & 69.72 & 77.77 & 68.60 & 93.43 & 53.98 & 86.67 & 75.58 \\
    & V-Net~\cite{milletari2016v} & 68.81  & - & 75.34 & 51.87 & 77.10 & 80.75 & 87.84 & 40.05 & 80.56 & 56.98 \\
    & Attention-UNet~\cite{schlemper2019attention} & 77.77 & 36.02 & 89.55 & 68.88 & 77.98 & 71.11 & 93.57 & 58.04 & 87.30 & 75.75 \\
    & nnUNet \cite{isensee2021nnunet} & 83.98 & 12.56 & 91.73 & 66.22 & \underline{87.30} & 84.41 & \underline{96.15} & 76.04 & 94.81 & 75.20 \\
    \midrule
    
    & ViT~\cite{dosovitskiy2020vit} & 71.29 & 32.87 & 73.73 & 55.13 & 75.80 & 72.20 & 91.51 & 45.99 & 81.99 & 73.95 \\
    
    & TransUNet \cite{chen2021transunet} & 77.48 & 31.69 & 87.23 & 63.13 & 81.87 & 77.02 & 94.08 & 55.86 & 85.08 & 75.62 \\
    
    & CoTr \cite{xie2021cotr} & 78.08 & 27.38 & 85.87 & 61.38 & 84.83 & 79.36 & 94.28 & 57.65 & 87.74 & 73.55 \\
    
    & UNETR \cite{hatamizadeh2022unetr} & 79.57 & 23.87 & 89.99 & 60.56 & 85.66 & 84.80 & 94.46 & 59.25 & 87.81 & 73.99 \\
    
    & SwinUNETR~\cite{tang2022swinunetr} & 77.09 & 34.21 & 88.73 & 60.07 & 78.10 & 80.99 & 93.81 & 51.97 & 89.72 & 73.29 \\
    
    & SwinUNet \cite{cao2021swinunet} & 79.13 & 21.55 & 85.47 & 66.53 & 83.28 & 79.61 & 94.29 & 56.58 & 90.66 & 76.60 \\
    
     & CTO-Net~\cite{lin2023rethinking} & 81.10 & 18.75 & 87.72 & 66.44 & 84.49 & 81.77 & 94.88 & 62.74 & 90.60 & 80.20 \\
    
    \multirow{-8}{*}{\rotatebox{90}{Transformer-based}} & D-LKA Net~\cite{azad2024beyond} & 83.37 & 16.52 & \textbf{91.86} & 68.91 & 86.17 & \underline{85.02} & 95.56 & 71.24 & 89.99 & 78.46 \\
    \midrule
    
    \multirow{5}{*}{\rotatebox{90}{MLP-based}}& Mixer~\cite{tolstikhin2021mixer} & 82.80 & 21.12 & 87.40 & 67.16 & 85.37 & 83.96 & 95.16 & 70.53 & 94.53 & 78.25 \\
    & UNeXt~\cite{valanarasu2022unext}  & 67.07 & 40.47 & 76.43 & 51.64 & 74.54 & 67.94 & 91.11 & 34.95 & 79.20 & 60.70 \\
    & CycleMLP \cite{chen2022cyclemlp} & 70.82 & 24.47 & 75.07 & 61.75 & 77.25 & 70.71 & 92.83 & 42.29 & 81.44 & 65.26  \\
    & Shift~\cite{Lian_2021_ASMLP} & 83.69 & 16.44 & 91.35 & 64.50 & 86.68 & 84.15 & 96.01 & 73.62 & 95.22 & 78.01 \\
    & Wave~\cite{tang2022wave} & \underline{85.09} & \underline{12.49} & \underline{91.74} & \underline{72.48} & 86.37 & 84.63 & 96.03 & \textbf{77.02} & \textbf{95.25} & 77.23 \\
    \midrule
    \multicolumn{2}{r|}{SAM~\cite{kirillov2023sam}} & 44.87 & 71.71 & 44.61 & 26.14 & 49.70 & 58.22 & 82.95 & 9.12 & 51.83 & 36.38 \\
    \multicolumn{2}{r|}{MedSAM-Fine-tune~\cite{MedSAM}} & 74.28 & 42.51 & 67.14 & 56.19 & 82.22 & 77.76 & 92.72 & 55.45  & 87.85 & 74.94 \\
    \multicolumn{2}{r|}{MedSAM-LoRA~\cite{zhang2023customized}} & 77.62 & 31.33 & 81.40 & 60.96 & 78.38 & 77.33 & 93.84 & 62.24 & 88.29 & \underline{78.50} \\
    \midrule
    
    \multicolumn{2}{r|}{PHNet (Ours)} & \textbf{85.62} & \textbf{11.75} & 90.31 & \textbf{74.08} & \textbf{87.46} & \textbf{85.69} & \textbf{96.16} & \underline{76.94} & \underline{95.23} & \textbf{79.11} \\
    \bottomrule
    \end{tabular}}
    \label{Synapse_sota}
    \end{table*}

\subsection{Implementation Details}
PHNet is implemented using PyTorch and MONAI~\cite{cardoso2022monai} framework and trained on an NVIDIA RTX 3090 GPU. 
For COVID-19-20, all images are interpolated into the voxel spacing of $ 0.74 \times 0.74 \times 5.00 \textup{mm}^3$. 
Three sub-volumes of $ 224 \times 224 \times 28$ are sampled from each scan. 
For LiTS, we perform random cropping of sub-volumes with the shape of $96 \times 96 \times 96$ from each case. 
By using fixed-sized patches, we ensure that a single architecture can be applied consistently to all data within
the same dataset. Across datasets, we have implemented an automatic configuration strategy to adjust the number of 2D conv-blocks and 3D conv-blocks in PHNet based on the input resolution of each dataset.
We set the max training epoch to 250 on Synapse, 450 on COVID-19-20, and 200 on both the LiTS and MSD BraTS.
The model with the best validation performance is selected for testing.
For all experiments, we adopt the AdamW optimizer~\cite{loshchilov2017decoupled} with an initial learning rate $lr = 10^{-3} \times \frac{\text{batch\_size}}{1024}$, as suggested by~\cite{hou2022vip}. 
The objective function is the summation of Dice loss and cross-entropy loss, which is a commonly employed approach in this domain.
Except for the above, we follow baselines from~\cite{roth2022rapid}, ~\cite{isensee2021nnunet}, ~\cite{zhang2022deep} and ~\cite{peiris2022robust} for the COVID-19-20, Synapse, LiTS and BraTS datasets, respectively.

\begin{table*}[t]
    \centering
    \caption{Result comparisons with the state-of-the-art methods and top-13 solutions on COVID-19-20~\cite{roth2022rapid}.}
    \renewcommand\arraystretch{1.2}
    \setlength{\tabcolsep}{15.5pt}{
    \begin{tabular}{c | r|ccccc cc}
    \toprule
    \multicolumn{2}{c|}{\textbf{Methods}} & \textbf{Dice} $\uparrow$ & \textbf{IoU} $\uparrow$ & \textbf{SD} $\uparrow$ & \textbf{NVD} $\downarrow$ & \textbf{HD} $\downarrow$ & \textbf{Method} & \textbf{Dice} $\uparrow$ \\
    \cmidrule(lr){1-7} \cmidrule(lr){8-9}
    \multirow{2}{*}{\rotatebox{90}{CNN}}& UNet~\cite{falk2019unet} & 69.09 & 55.27 & 62.83 & 40.79 & 134.76 & Rank 1 & 77.09 \\
    & nnUNet \cite{isensee2021nnunet} & 72.51 & 59.40 & 69.04 & 27.87 & 123.65 & Rank 2 & 76.87  \\ 
    \cmidrule(lr){1-7} 
    & TransUNet~\cite{chen2021transunet}  & 18.64 & 11.70 & 11.93 & 65.06 & 279.03 & Rank 3 & 76.87  \\
    
    & CoTr~\cite{xie2021cotr}             & 59.63 & 45.65 & 52.97 & 44.57 & 172.78 & Rank 4 & 76.78 \\
    
    & UNETR \cite{hatamizadeh2022unetr} & 57.18 & 43.10 & 51.20 & 44.64 & 174.40 &  Rank 5 & 76.77  \\

    & SwinUNETR~\cite{tang2022swinunetr}  & 63.65 & 50.13 & 58.19 & 37.20 & 141.42 & Rank 6 & 76.64  \\
    
    & SwinUNet~\cite{cao2021swinunet}   & 32.82 & 22.08 & 20.71 & 67.62 & 221.46 & Rank 7 & 76.47  \\
    
    \multirow{-6}{*}{\rotatebox{90}{{Transformer}}} & D-LKA Net~\cite{azad2024beyond}& 66.11& 53.32& 62.64& 38.93& 136.67& Rank 8 & 76.45 \\
    
    \cmidrule(lr){1-7} 
    
    \multirow{4}{*}{\rotatebox{90}{MLP}} &
    Shift~\cite{Lian_2021_ASMLP} & 72.18 & 59.32 & 68.58 & 25.27 & 132.01 & Rank 9 & 76.28 \\ &
    
    Wave~\cite{tang2022wave} & \underline{74.08} & \underline{60.92} & \underline{70.76} & \underline{22.12} & \underline{120.59} & Rank 10 & 76.27 \\ &   
    
    UNeXt~\cite{valanarasu2022unext}    & 21.90 & 14.64 & 10.79 & 81.52 & 328.83 & Rank 11 & 76.27  \\ 
    
    & CycleMLP~\cite{chen2022cyclemlp}    & 27.53 & 17.91 & 15.55 &  79.67 & 228.07  & Rank 12 & 76.16  \\

    \cmidrule(lr){1-7} 
    \multicolumn{2}{r|}{MedSAM-LoRA~\cite{zhang2023customized}}    & 21.11 & 13.97 & 10.93 & 53.11 & 290.30 & Rank 13 & 76.06 \\
    \cmidrule(lr){1-7} \cmidrule(lr){8-9}
    
    \multicolumn{2}{r|}{PHNet (Ours)} & \textbf{76.34} & \textbf{63.36} & \textbf{72.10} & \textbf{20.60} & \textbf{108.16} & Ours & \textbf{77.18}  \\
    \bottomrule
    \end{tabular}}
    \label{COVID-19-20_sota}
\end{table*}

\begin{table*}[thbp]
    \centering
    \caption{Result comparisons with state-of-the-art methods on LiTS~\cite{bilic2023liver} and MSD BraTS~\cite{antonelli2022medical}.}
    \renewcommand\arraystretch{1.2}
    \setlength{\tabcolsep}{7pt}{
    \begin{tabular}{ c| c c c |cc cc c c c c c}
    \multicolumn{4}{c}{LiTS} & \multicolumn{9}{c}{MSD BraTS} \\
    \toprule
    \multicolumn{2}{c}{\multirow{2}{*}{Methods}} & \multirow{2}{*}{Dice$\uparrow$} & \multirow{2}{*}{IoU$\uparrow$} & \multirow{2}{*}{Methods} & \multicolumn{2}{c}{Mean} & \multicolumn{2}{c}{WT} & \multicolumn{2}{c}{ET} & \multicolumn{2}{c}{TC} \\
    \cmidrule(r){6-7}\cmidrule(r){8-9}\cmidrule(r){10-11}\cmidrule(r){12-13}
    \multicolumn{2}{c}{}&&&& HD$\downarrow$ & Dice$\uparrow$ & HD$\downarrow$ & Dice$\uparrow$ & HD$\downarrow$ & Dice$\uparrow$ & HD$\downarrow$ & Dice$\uparrow$  \\
    \midrule
    \multirow{3}{*}{\rotatebox{90}{\small{CNN}}} &
    UNet~\cite{ronneberger2015u} & 85.38 & 75.22 & UNet~\cite{ronneberger2015u} & 5.13 & 83.58 & 3.5 & 91.28 & 7.21 & 75.86 & 4.65 & 83.59 \\
    
    & Attention-UNet~\cite{schlemper2019attention} & 89.98 & 82.13 & Attention-UNet~\cite{schlemper2019attention} & 9.97 & 66.4 & 9 &	76.7 & 10.45 & 54.3 & 10.46 & 68.3 \\
    
    & VNet~\cite{milletari2016v} & 89.72 & 80.71 & nnUNet~\cite{isensee2021nnunet} & 4.2 & \underline{86.07} & 3.64 & \textbf{91.9} & 4.06 & 80.97 & 4.91 & 85.35 \\
    \midrule

    & ViT~\cite{dosovitskiy2020vit} & 83.67 & 72.49 & TransBTS~\cite{wang2021transbts} & 9.65 & 69.6 & 10.03 & 77.9 & 9.97 & 57.4 & 8.95 & 73.5 \\
    
    & TransUNet~\cite{chen2021transunet} & 82.4 & 70.62 & TransUNet~\cite{chen2021transunet} & 12.98 & 64.4 & 14.03 & 70.6 & 10.42 & 54.2 &	14.5 & 68.4 \\
    
    & SwinUNet~\cite{cao2021swinunet} & 86.08 & 75.8 & CoTr~\cite{xie2021cotr} & 9.7 & 68.3 & 9.2 & 74.6	& 9.45 & 55.7 & 10.45 & 74.8 \\
    
    & UNETR~\cite{hatamizadeh2022unetr} & 81.48 & 69.04 & UNETR~\cite{hatamizadeh2022unetr} & 3.88 & 84.89 & 3.91 & 91.05 & \underline{3.05} & 79.14 & 4.68 & 84.47 \\
    
    & SwinUNETR~\cite{tang2022swinunetr} & 91.72 & 85.1 & SwinUNETR~\cite{tang2022swinunetr} & 4.16 & 85.06 & 3.61 & \underline{91.73} & 4.76 & 78.21 & \textbf{4.08} & 85.23 \\
    
    & D-LKA Net~\cite{azad2024beyond} & \underline{92.05} & \underline{85.54} & D-LKA Net~\cite{azad2024beyond} & 3.90 & 85.09 & \textbf{3.54} & \underline{91.73} & \textbf{2.04} & \textbf{83.67} & 6.12 & 79.89 \\
    
    \multirow{-6}{*}{\rotatebox{90}{\small{Transformer}}} &
    CTO-Net~\cite{lin2023rethinking} & 91.5 & 84.59 & VT-UNet~\cite{peiris2022robust} & 3.84 & 85.9 & 4.01 & 90.8 & \underline{2.91} & 81.8 & \underline{4.49} & 85.0 \\
    \midrule

    \multirow{3}{*}{\rotatebox{90}{\small{MLP}}} &
    UNeXt~\cite{valanarasu2022unext} & 84.37 & 73.15 & Mixer~\cite{tolstikhin2021mixer} & 4.56 & 85.28 & 3.69 & 91.47 & 4.88 & 80.22 & 5.12 & 84.1 \\
    
    & CycleMLP~\cite{chen2022cyclemlp} & 88.27 & 79.34 & Shift~\cite{Lian_2021_ASMLP} & \underline{3.8} & 85.81 & 3.68 & 91.35 & 3.12 & 81.4 & 4.6 &	84.69 \\
    
    & Wave~\cite{tang2022wave} & 91.68 & 84.75 & Wave~\cite{tang2022wave} & 4.02 & 85.65 & 3.64 & 91.56 & 3.90 &	79.5 & 4.51 & \textbf{85.89} \\
    \midrule
    
   \multicolumn{2}{c}{PHNet(Ours)} & \textbf{92.17} & \textbf{85.68} & Ours & \textbf{3.76} & \textbf{86.62} & \underline{3.57} & 91.41 & 3.2 & \underline{82.86} & 4.51 & \underline{85.59} \\
    \bottomrule
    \end{tabular}
    }
    \label{tab:lits_brats_sota}
\end{table*}

\subsection{Comparisons with State-of-the-Arts} 
\subsubsection{Comparison methods.} 
We compare the proposed PHNet with CNN-based, Transformer-based, MLP-based methods, and foundation models.
We list the details below. 
\begin{itemize}
\item\textit{CNN-based methods}, including V-Net~\cite{milletari2016v}, UNet~\cite{ronneberger2015u}, nnUNet~\cite{isensee2021nnunet}, and Attention-UNet~\cite{schlemper2019attention}. 
These architectures feature a hierarchical contracting path for context aggregation and a symmetric expanding path for resolution recovery and precise localization.
\item\textit{Transformer-based methods}, including ViT~\cite{dosovitskiy2020vit}, UNETR~\cite{hatamizadeh2022unetr}, SwinUNETR~\cite{tang2022swinunetr}, TransUNet~\cite{chen2021transunet}, SwinUNet~\cite{cao2021swinunet}, CoTr~\cite{xie2021cotr}, CTO-Net~\cite{lin2023rethinking},VT-UNet~\cite{peiris2022robust} and TransBTS~\cite{wang2021transbts}.
These architectures can be categorized into three groups: 1) classical Transformer (\ie, ViT), 2) encoder-decoder framework utilizing pure Transformer blocks, including SwinUNet and VT-UNet, and 3) hybrid architectures that combine CNN and Transformer, including UNETR, SwinUNETR, TransUNet, CoTr, CTO-Net and TransBTS.
\item\textit{MLP-based methods}, including UNext~\cite{valanarasu2022unext}, CycleMLP~\cite{chen2022cyclemlp}, MLP-Mixer (Mixer)~\cite{tolstikhin2021mixer}, ShiftMLP (Shift)~\cite{Lian_2021_ASMLP}, and WaveMLP (Wave)~\cite{tang2022wave}.
We only replace the MLPP module in PHNet with these alternatives for a fair comparison.

\item\textit{MedSAM}~\cite{MedSAM} builds upon SAM~\cite{kirillov2023sam}, a foundation model for semantic segmentation.
MedSAM retains the same architecture as SAM, which features a ViT-based encoder, a prompt module, and a mask decoder. To adapt MedSAM for target tasks, we employ two fine-tuning strategies: Full model fine-tuning (MedSAM-Fine-tune) and LoRA~\cite{hu2021lora} fine-tuning (MedSAM-LoRA~\cite{zhang2023customized}).
\end{itemize}

\begin{figure*}[!t]
\centering
\includegraphics[width=0.99\textwidth]{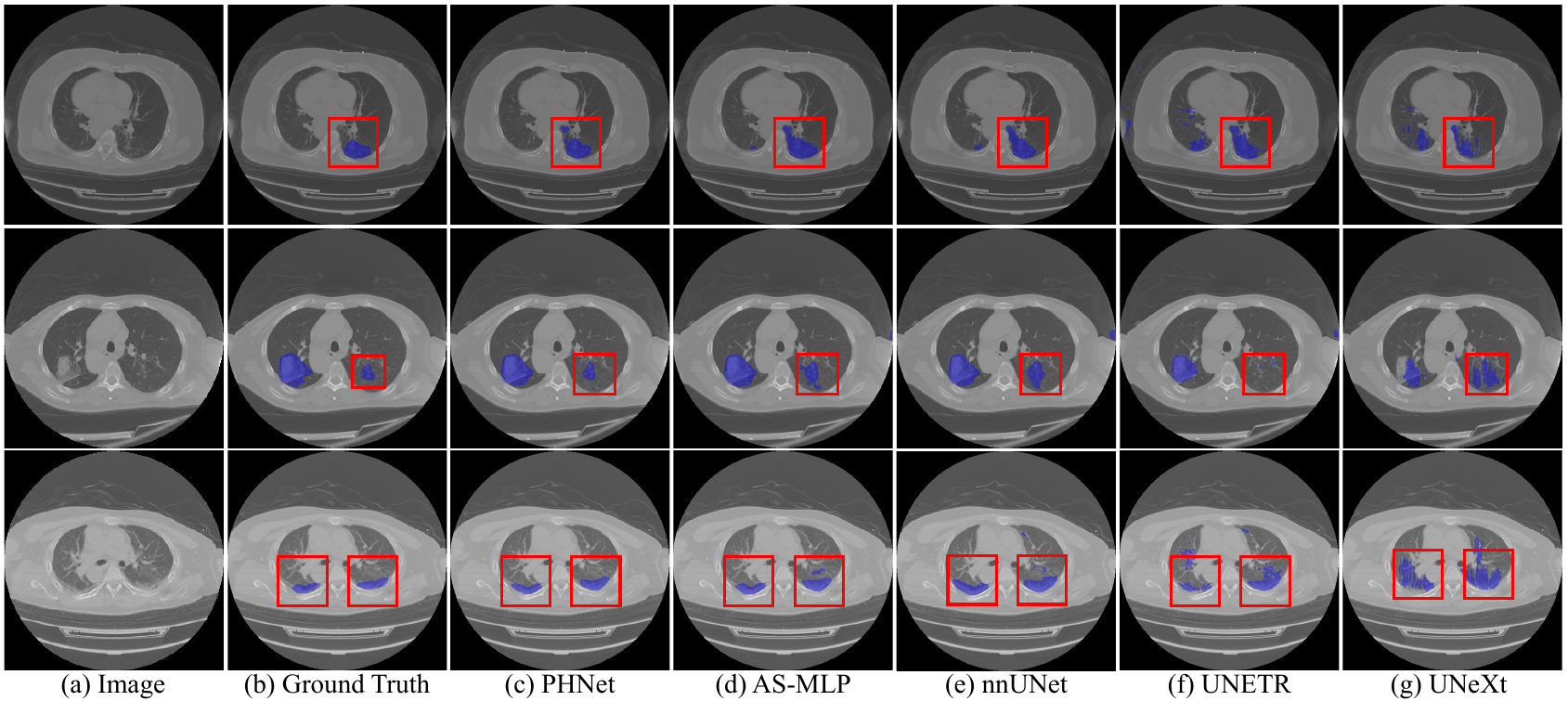}
\caption{Qualitative visualizations of different methods on COVID-19-20. The red boxes highlight the main differences.}
\label{fig:covid_visual}
\end{figure*}
\begin{figure*}[t]
\centering
\includegraphics[width=0.99\textwidth]{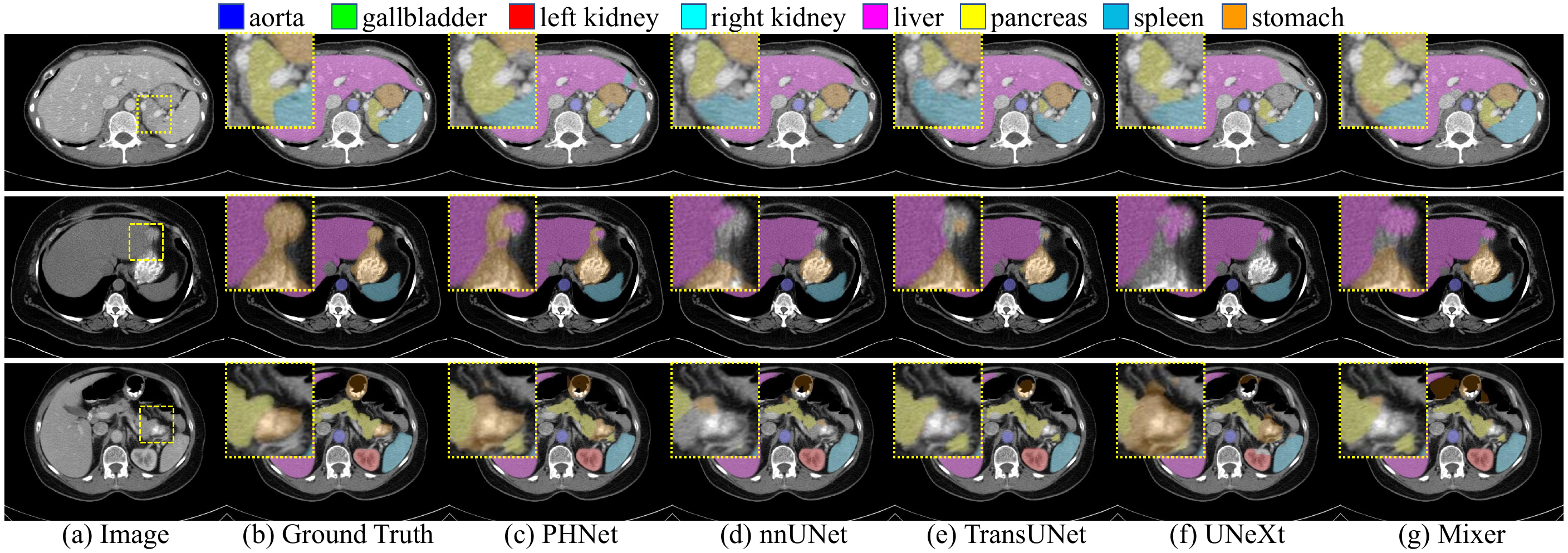}
\caption{Qualitative visualizations of different methods on the Synapse~\cite{landman2015miccai}. Regions of evident improvements are enlarged to show better details.}
\label{fig:visual}
\end{figure*}

\subsubsection{Quantitative comparisons.} 
We begin by evaluating the performance of our method in multi-organ segmentation on Synapse~\cite{landman2015miccai} in Table~\ref{Synapse_sota}. 
Result shows that our method achieves the highest average Dice score of 85.62$\%$ and lowest HD of 11.75, outperforming the SOTA methods. 
Specifically, our method achieves 1.64$\%$ and 8.77$\%$ Dice improvements over nnUNet~\cite{isensee2021nnunet} and UNet~\cite{falk2019unet}, whose backbones are built upon 3D CNNs. 
Compared to top-performing competitors of Transformer-based CTO-Net~\cite{lin2023rethinking}, UNETR~\cite{hatamizadeh2022unetr}, MLP-based UNext~\cite{valanarasu2022unext}, Wave~\cite{tang2022wave}, our method also achieves better performance with 4.52$\%$, 6.05$\%$, 18.55$\%$, 0.53$\%$ Dice gains, respectively. 
These results conform to our argument that PHNet obtains satisfying segmentation performance through effective local-to-global modeling.
The segmentation performance of pancreas is relatively lower than other organs. It is due to the low contrast between the pancreas and surrounding tissues. The pancreas is a small organ with a complex shape and is located near other organs with similar intensity values. The pancreas is also subject to various deformations due to respiration and other physiological processes. These factors make it difficult to accurately segment the pancreas using common segmentation networks.
Compared with pretrained vision encoders like SAM and MedSAM, we found that directly applying pretrained SAM model to the target dataset would achieve unsatisfactory performance, which achieves 44.87\% in Dice score.
Fine-tuning or using parameter-efficient fine-tuning like LoRA to the universal vision encoder on the small dataset can significantly improve the performance, but it is still worse than the specialized model.

We further evaluate the performance on COVID-19-20~\cite{roth2022rapid} and the official evaluation result is presented in Table~\ref{COVID-19-20_sota}. 
Compared to CNN-based methods, our method attains the highest scores in all metrics. This suggests that CNN-based approaches have limitations in long-distance context fusion.
Compared to Transformer-based methods, PHNet also achieves better performance. This is because Transformer-based methods commonly divide the input image into patches to fit GPU memory constraints, which may overlook local features, particularly in the context of small lesions.
Compared to MLP-based methods, PHNet outperforms by a remarkable margin. 
This is partly because existing MLP-based methods are designed to segment on 2D slices which lose through-plane features, resulting in severe performance deterioration in challenging volumetric image segmentation tasks. 
The large performance gap between MedSAM-LoRA~\cite{zhang2023customized} and PHNet indicates the constrained generalization capacity of foundation models within specific segmentation tasks.
Additionally, following~\cite{roth2022rapid}, we perform five-fold cross-validation and model ensemble using our proposed method.
The result demonstrates that our method achieves the highest dice score of 77.18\%, surpassing the top-12 solutions in this challenge\footnote{\url{https://covid-segmentation.grand-challenge.org/evaluation/challenge/leaderboard}}.

Additionally, the performance evaluation on the LiTS and MSD BraTS datasets is presented in Table~\ref{tab:lits_brats_sota}. 
Our proposed PHNet demonstrates superior results, achieving the highest Dice score of 92.17\% and IoU of 85.68\% on the LiTS dataset. 
This surpasses the performance of the best competitors of CNN-based methods, Transformer-based methods and MLP-based methods, including Attention-UNet~\cite{schlemper2019attention}, SwinUNETR~\cite{tang2022swinunetr}, and WaveMLP~\cite{tang2022wave}. 
Notably, our method exhibits Dice gains of 2.19\%, 0.45\%, and 0.49\%, and IoU gains of 3.55\%, 0.58\%, and 0.93\% over these respective competitors. 
On the MSD BraTS dataset, our PHNet achieves the highest average Dice score of 86.62\% and the lowest average HD of 3.76. Once again, it outperforms the top competitors, namely nnUNet~\cite{isensee2021nnunet}, SwinUNETR~\cite{tang2022swinunetr}, and Shift~\cite{Lian_2021_ASMLP}. 
These results provide further evidence to support our claim that PHNet achieves outstanding segmentation performance through effectively incorporating local-to-global modeling.
It is worth highlighting that BraTS encompasses isotropic data, underscoring the resilience of our approach across both anisotropic and isotropic datasets.

\subsubsection{Qualitative comparisons.}
Visual comparisons on the Synapse~\cite{landman2015miccai} dataset are presented in Figure~\ref{fig:visual}. 
Among the compared methods, PHNet demonstrates the best visual segmentation results when compared to nnUNet~\cite{isensee2021nnunet}, TransUNet~\cite{chen2021transunet}, UNext~\cite{valanarasu2022unext}, and Mixer~\cite{tolstikhin2021mixer}.
As shown In the first row, it is evident that our method achieves superior segmentation, addressing the issue of under-segmentation observed in nnUNet, TransUNet, and UNext. 
Furthermore, in the second row, PHNet accurately distinguishes boundaries between the liver and stomach, while both UNeXt and Mixer exhibit over-segmentation of the liver. 
Likewise, in the last row, PHNet provides a more precise delineation between the pancreas and stomach. We also present visual comparisons on the COVID-19-20 dataset in Figure~\ref{fig:covid_visual}. We compare the segmentation outcomes of our proposed PHNet with several state-of-the-art methods, namely AS-MLP~\cite{Lian_2021_ASMLP}, nnUNet~\cite{isensee2021nnunet}, UNETR~\cite{hatamizadeh2022unetr}, and UNeXt~\cite{valanarasu2022unext}. Upon examination, it becomes evident that our PHNet achieves the most exceptional segmentation results, characterized by both semantic consistency and boundary integrity. In contrast, the segmentation results obtained by the other methods appear either incomplete or inaccurate.
These results can be attributed to the effective local-to-global modeling and aggregation of inter-axis and intra-axis token features in PHNet, which facilitates the extraction of accurate contours. 
Overall, PHNet excels in segmentation, alleviating the challenges associated with under- or over-segmentation of contours.
Drawing upon the encoder-decoder architecture, PHNet exhibits promise for generalization to diverse medical image segmentation tasks~\cite{lin2019automated,hering2022learn2reg,dayarathna2024deep}, as evidenced by its consistent performance across various datasets.

\subsection{Ablation Study} 
We conduct ablation studies to validate the effectiveness and efficiency of each component in our method.
We use the same decoder architecture for all variations.

\begin{figure*}[!htbp]
\centering
\includegraphics[width=0.85\textwidth]
{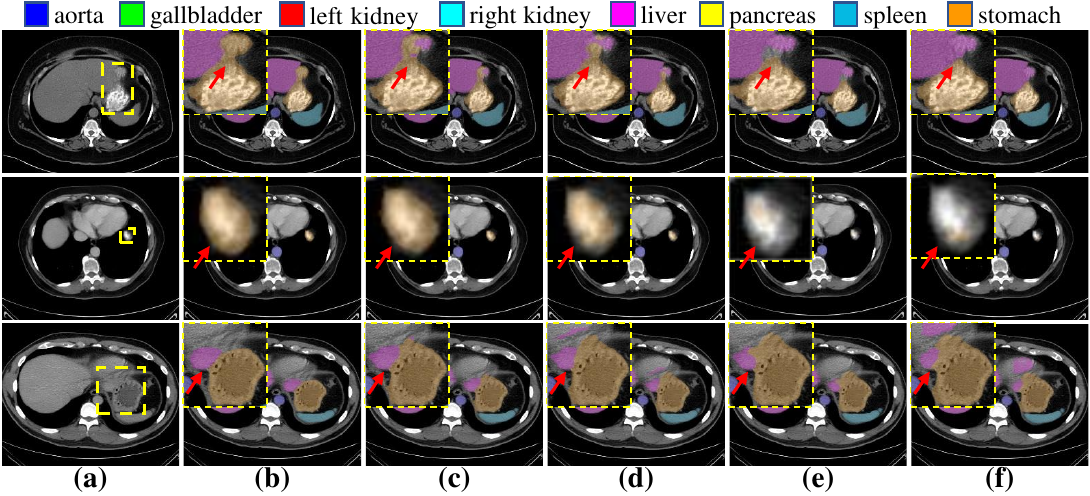}
\caption{Segmentation visualizations of ablation variations on Synapse~\cite{landman2015miccai}, including (a) original image, (b) ground truth, predictions of (c) {baseline} + {IP} + {TP} + {AA}, (d)  {baseline} +  {IP} +  {TP}, (e)  {baseline} +  {IP}, and (f)  {baseline}. Regions of evident improvements are enlarged to show better details. Better viewed with zooming in.}
\label{fig:visual_abl_Synapse}
\end{figure*}
\begin{table}[t]
\centering
\caption{Result comparisons with different effects of each component on Synapse~\cite{landman2015miccai}. ``{Base}'' denotes the vanilla MLP network. ``{IP-}''denotes the proposed axial decomposition operation in IP-MLP module. ``{TP-}'' denotes the proposed operation of decomposition of in-plane feature and through-plane feature in TP-MLP. ``{AA-}'' denotes our proposed AA-MLP module.}
\renewcommand\arraystretch{1.2}
\setlength{\tabcolsep}{7pt}{
\begin{tabular}{c c c c|c c|c c}
     \toprule
     \textbf{Base} &\textbf{IP-} & \textbf{TP-} & \textbf{AA-} & \textbf{Dice} $\uparrow$ & \textbf{HD} $\downarrow$ & \textbf{FLOPs} & \textbf{Thro.} \\
     \midrule
     \cmark &       &       &       & 82.80 & 21.12 & 1249 & 1.04  \\
     \cmark &\cmark &       &       & 84.29 & 19.05 & 938 & 1.73  \\
     \cmark &      &\cmark & & 83.39 & 19.66 & 971 & 1.62 \\
     \cmark &\cmark &\cmark &       & 84.94 & 18.29 & 947 & 1.75  \\
     \cmark &\cmark &\cmark &\cmark & \textbf{85.62} & \textbf{11.75} & 953 & 1.73 \\
     \bottomrule
     \end{tabular}}
    \label{tab:ablation_Synapse}
 \end{table}
\begin{table}[t]
    \centering
    \caption{Comparisons with interactions of different axes. W, H, D denotes horizontal, vertical and depth axis, respectively.}
    \renewcommand\arraystretch{1.2}
    \setlength{\tabcolsep}{6pt}{
    \begin{tabular}{cccccc}
    \toprule
    & WH (Ours) & WH+WD & WH+HD & WH+WD+HD \\
    \midrule
    Dice/HD & \textbf{85.62}/\textbf{11.75} & 84.38/14.63 & 84.58/12.52 & 84.62/13.15 \\
    \bottomrule
    \end{tabular}}
    \label{tab:ablation_TP}
\end{table}

\subsubsection{Effectiveness of each components.}
To verify the effectiveness of core components in our approach, we increase each essential component gradually based on the vanilla MLP network (abbreviated as ``Base''). We conduct experiments on the Synapse dataset, as shown in Table~\ref{tab:ablation_Synapse}.
Compared with the baseline, the integration of axial decomposition yields a 1.49\% enhancement in Dice and a noteworthy improvement in efficiency. 
This improvement can be attributed to the encoding of positional information and the reduction in computational complexity.
There is an additional improvement of 0.65\% Dice through decomposition into in-plane (IP) and through-plane (TP) features and training them separately in a sequential manner. 
This design choice aligns with the reading process of radiologists, who primarily examine IP slices and utilize TP information as supplementary.
Furthermore, AA-MLP contributes to 0.68\% performance gains with slightly increasing computational cost, and the whole framework achieves an 85.62\% Dice score. 
This is because intra-axis token communication enables enhanced learning representation from larger receptive field, as shown in Figure~\ref{fig:visual_abl_Synapse}. On the other hand, the performance boost indicates that partial interactions are
complementary to content features and compact with less irrelevant information.
We delve deeper into exploring the interactions between different axes in our research. 
Experiment results in Table~\ref{tab:ablation_TP} reveal some interesting findings. 
Specifically, we observe that incorporating only the in-plane (IP) cross-axial interaction (WH) with AA-MLP leads to a significant improvement in performance. 
However, the introduction of through-plane (TP) cross-axial interactions (WD and HD) may introduce redundant information, resulting in a degradation of performance. 
A similar trend can be observed in the evaluation of HD metric, further underscoring the effectiveness of each component in our approach.

\begin{table}[!htbp]
      \centering
    \caption{Comparison between various number of 2D and 3D CNN layers on Synapse dataset.}
    \resizebox{0.5\textwidth}{!}{\renewcommand\arraystretch{1.2}
      \begin{tabular}{ c c c c c | c c c c} 
          \toprule
          & \multicolumn{4}{c|}{2D CNN} & \multicolumn{4}{c}{3D CNN} \\
          \midrule
          & 1 & 2 & 4 & 8 & 1 & 2 & 4 & 8 \\
          \midrule
          Dice & 82.31 & 85.62 & 84.72 & 82.89 & 85.62 & 83.42 & 85.09 & 84.11 \\
          HD & 18.64 & 11.75 & 12.23 & 17.30 & 11.75 & 16.43 & 13.66 &  16.57 \\
          \bottomrule
          \end{tabular}}
\label{tab:conv_ablation_Synapse}
\end{table}

\begin{figure}[htbp]
\centering
\includegraphics[width=0.48\textwidth]
{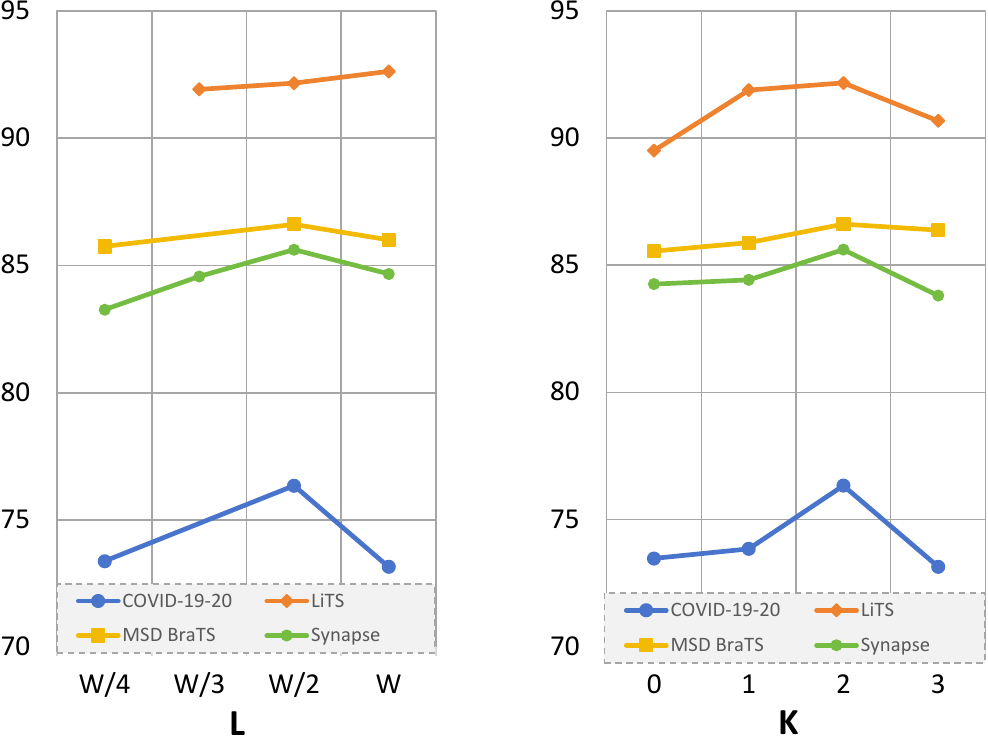}
\caption{Impact of (a) different segment length $L$; (b) different number of MLPP layers $K$ on COVID-19-20, Synapse and MSD BraTS datasets.}
\label{fig:ablation}
\end{figure}

\subsubsection{Impact of CNN layers.}
In Table~\ref{tab:conv_ablation_Synapse}, we evaluate the number of 2D and 3D convolutional layers. We conduct experiments on the Synapse dataset. The results indicate that the utilization of two 2D convolutional layers yields the best outcome. This can be attributed to the fact that the in-plane resolution of images in the Synapse dataset is roughly $\frac{1}{4}$ of the through-plane resolution.  By employing two 2D CNN layers, the feature is reformulated in approximately uniform resolution across
all three axes. In the case of 3D convolutional layers, experimental results reveal that there are no significant performance differences when varying the number of layers. The consistent performance across different layer numbers highlights robustness of our proposed method.

\subsubsection{Impact of segment length.}
In Figure~\ref{fig:ablation}(a), we investigate the impact of different segment lengths $L$ in PHNet. We conduct experiments on COVID-19-20, Synapse, LiTS, and MSD BraTS datasets. Expressly, the segment length is set to various ratios of the width ($W$), \ie, $1$, $\frac{1}{2}$, $\frac{1}{3}$, and $\frac{1}{4}$, respectively. With a larger segment length, long-range dependencies can be more effectively captured in deep layers. Conversely, smaller segment length implies fewer adjacent tokens are grouped, emphasizing more on local information. Remarkably, setting $L = W/2$ consistently yields the highest Dice score across Synapse, COVID-19-20, and MSD BraTS datasets. However, for the LiTS dataset, the best performance is achieved when setting $L = W$, which aligns with our expectations as the liver organ typically exhibits a larger size. These results effectively demonstrate the effectiveness of token segmentation and the sufficiency of partial interactions in diverse applications.

\subsubsection{Impact of MLP layers.}
In Figure~\ref{fig:ablation}(b), we study the influence of a different number of MLP layers $K$ in our PHNet. We conduct experiments on COVID-19-20, Synapse, LiTS, and MSD BraTS datasets. The optimal number of MLP layers slightly varies across different datasets, indicating that the number of MLP layers may be dataset-specific. Nevertheless, adding MLP layers consistently improves performance across all datasets, highlighting the effectiveness of utilizing MLP layers to capture long-range dependencies.
\begin{figure}[htbp]
\centering
\includegraphics[width=0.46\textwidth]{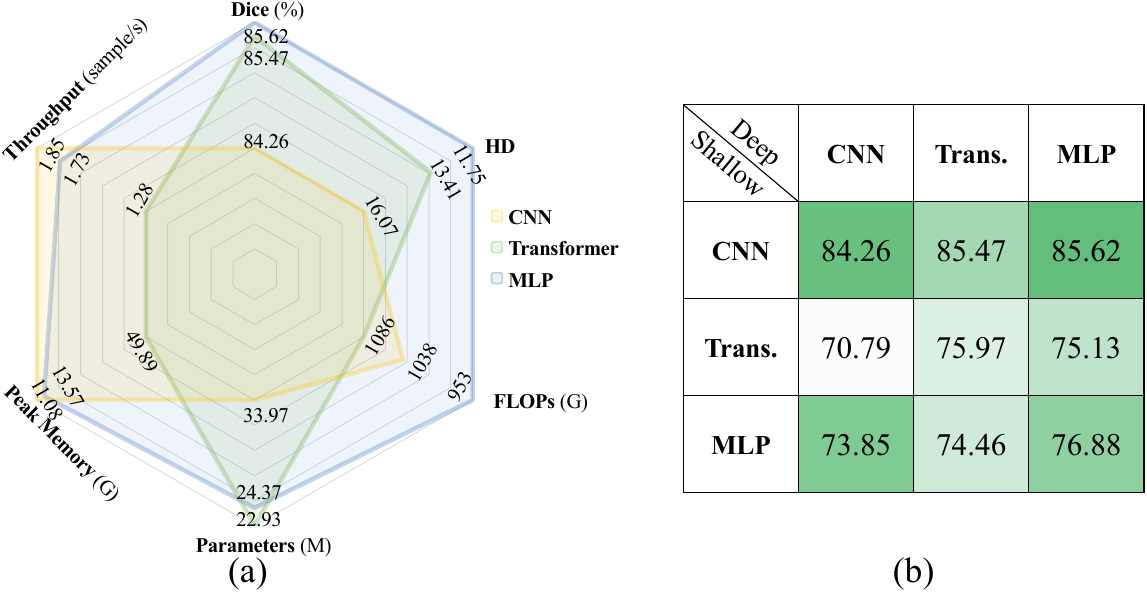}
\caption{(a) Comparisons between conventional architectures in terms of effectiveness and efficiency; (b) Comparison between architecture combinations in Dice (\%).}
\label{fig:discussion}
\end{figure}

\begin{table}[htbp]
     \caption{Comparison between different model capacities of CNN, Transformer, and MLPP on Synapse~\cite{landman2015miccai} in terms of effectiveness. S, B, L depends on the number of blocks in each layer and denotes small, base, and large, respectively.}
     \renewcommand\arraystretch{1.2}
    \setlength{\tabcolsep}{12.5pt}{
     \begin{tabular}{ccc c c }
     \toprule
     
    {\textbf{Methods}} & {\textbf{Block}} 
     & {\textbf{Dice} $\uparrow$}
     & {\textbf{FLOPs}} & {\textbf{Params}}   \\
     \midrule
        
     CNN-S & (2, 2) & 79.83 & 864 & 15.02  \\

     CNN-B & (6, 2) & 82.04 & 914 & 20.01 \\

     CNN-L & (8, 4) & 83.51 & 955 & 27.96 \\
     \midrule
     
     Trans-S & (3, 2) & 80.56 & 875 & 11.43 \\

     Trans-B & (6, 3) & 82.62 & 919 & 15.61 \\

     Trans-L & (9, 4) & 84.27 & 963 & 18.39 \\
    \midrule
    
     MLPP-S & (5, 2) & 81.16 & 870 & 14.16 \\

     MLPP-B & (10, 3) & 83.49 & 912 & 19.26 \\

     MLPP-L & (15, 4) & 85.62 & 953 & 24.37 \\
     
     \bottomrule     
     \end{tabular}}
    \label{tab:synapse accuracy comparison}
\end{table}

\subsection{Discussion}
In this section, We conduct three experiments to assess the efficiency and effectiveness of different backbone networks, including CNN, Transformer and our MLPP module, to validate our analysis. In all experiments, for the CNN component, we employ 3D CNN as it has exhibited superior performance compared to 2D CNN in the context of volumetric medical image segmentation tasks. As for the Transformer component, we utilize the Swin Transformer block~\cite{tang2022swinunetr} due to the promising performance and notable reduction in computational cost through the use of shift windows.

\subsubsection{Comparisons with conventional architectures.}
To evaluate the impact on efficiency and effectiveness, we replace the MLPP modules in the last two layers with an equal number of CNN and Transformer components, as depicted in Fig.~\ref{fig:discussion}(a). 
In comparison to CNN, our method achieves a significant 1.36\% improvement in Dice score, indicating that our proposed MLPP module more effectively encodes global information, resulting in enhanced segmentation accuracy.
Regarding computational efficiency, our method demonstrates a profile of 953G FLOPs, 24.37M parameters, 13.57G peak memory usage, and a throughput of 1.73 samples/second.
These efficiency outcomes are comparable to those of 3D CNN, indicating that our MLPP augments CNN's performance while upholding efficiency.
While Transformer showcases competitive performance when configured with stacked blocks, our method shows significant superiority in terms of efficiency. 
This suggests that our MLPP is capable of capturing global information while eliminating the exhaustiveness of token comparison seen in self-attention.

\subsubsection{Comparisons with architecture combinations.} 
We further explored different combinations of CNN, Transformer, and MLP across both shallow and deep layers of the encoder. 
The best performance was achieved by using CNN in shallow layers and MLP in deep layers, resulting in an impressive Dice score of 85.62\%. 
This supports our argument that CNN excels in capturing local features, while MLP is more effective in modeling long-range dependencies. 
Interestingly, the inverse configuration led to a substantial performance decline, indicating the importance of layer-specific functionality. We also observe that Transformer architectures exhibit underwhelming performance, possibly due to their high model complexity and the risk of overfitting on small datasets.

\subsubsection{Comparisons with architecture capacities.} 
To control computational complexity and evaluate effectiveness, we replace the last two layers of MLPP modules with CNN and Transformer components, as shown in Table~\ref{tab:synapse accuracy comparison}. 
We provide three variations of each comparison method by adjusting the number of blocks in each layer. 
In terms of Dice scores, the MLPP-based models outperform the CNN and Transformer variants with ratios of 81.16\% vs. 80.56\% vs. 79.83\%, 83.49\% vs. 82.62\% vs. 82.04\%, and 85.62\% vs. 84.27\% vs. 83.51\%. 
The results indicate that when considering similar computational complexity, MLPP-based models with a higher number of stacked blocks achieve superior Dice scores compared to the CNN and Transformer variants.
\section{Conclusion}
This paper introduced a novel permutable hybrid network, referred to as PHNet, specifically designed for volumetric medical image segmentation. 
By integrating 2D CNN, 3D CNN, and MLP, PHNet effectively captures both local and global features. 
Additionally, we proposed a permutable MLP block to address spatial information loss and alleviate computational burden. 
Experimental results on four public datasets demonstrate the superiority of PHNet over state-of-the-art approaches. 
Future research will explore extending the framework to other medical image analysis tasks, such as disease diagnosis and  localization, and further examine the interactions and effectiveness of CNN, Transformer, and MLP.
\bibliographystyle{IEEEtran}
\bibliography{refs}
\end{document}